\documentclass[twocolumn]{article}

\usepackage{ascmac, graphicx}

\title{\bf Secure Positioning of Mobile Terminals with Simplex Radio Communication}
\author{Mikio Fujii\\
{\normalsize Toshiba Solutions Corporation,}\\
{\normalsize 3-22 Katamachi, Fuchu-shi, Tokyo, 183-8512, JAPAN}\\
{\tt\normalsize Fujii.Mikio@toshiba-sol.co.jp}}

\date{}

\begin{document}

\maketitle

\begin{abstract}
With the rapid spread of various mobile terminals in our society, the importance of secure positioning is growing for wireless networks in adversarial settings. Recently, several authors have proposed a secure positioning mechanism of mobile terminals which is based on the geometric property of wireless node placement, and on the postulate of modern physics that a propagation speed of information never exceeds the velocity of light. In particular, they utilize the measurements of the round-trip time of radio signal propagation and bidirectional communication for variants of the challenge-and-response. In this paper, we propose a novel means to construct the above mechanism by use of {\it unidirectional} communication instead of bidirectional communication. Our proposal is based on the assumption that a mobile terminal incorporates a high-precision inner clock in a tamper-resistant protected area. In positioning, the mobile terminal uses its inner clock and the time and location information broadcasted by radio from trusted stations. Our proposal has a major advantage in protecting the location privacy of mobile terminal users, because the mobile terminal need not provide any information to the trusted stations through positioning procedures. Besides, our proposal is free from the positioning error due to claimant's processing-time fluctuations in the challenge-and-response, and is well-suited for mobile terminals in the open air, or on the move at high speed, in terms of practical usage. We analyze the security, the functionality, and the feasibility of our proposal in comparison to previous proposals.
\end{abstract}

\section{Introduction}

In the past decade, we have witnessed the successive emergence of various mobile terminals including mobile-phones, PDAs, handheld gaming devices, non-contact IC cards, RFID tags, and GPS receivers. They have pervaded and dramatically changed every aspect of our daily life in such a short time. As the mobile terminals became widespread, manufacturers made great efforts to meet urgent requirements of the market needs, and have made outstanding progress in key hardware technologies such as miniaturization of embedded components, lifetime extension of batteries, and sensitivity improvement of receivers. 

Today, the most popular wireless positioning system is perhaps the {\it civilian} GPS service, which is originally designed to provide location information from trusted satellites to exposed receivers in nonadversarial settings. Because all positioning procedures are presupposed to be legitimate by honest entities, the civilian GPS service has intrinsic vulnerabilities even to the most common attacks known as the impersonation attack, the modification attack, or the replay attack. In contrast, the {\it military} GPS service is secure against the impersonation attack and the modification attack by the external adversaries, thanks to encryption of GPS signals. But the service is only available to the United States military with their secret keys, and moreover, even the military GPS service is not secure enough to defend against the replay attack when it comes to location authentication.

A present RFID system also has security vulnerabilities on identification and location authentication especially to the replay attack, though expected to be a powerful tool for the product and commodity management. 
It is desirable for RFID tags to be equipped with a reliable security function for location authentication in the light of application demands to ensure the traceability of RFID tags in logistics and transportation systems.
It is no exaggeration to say that all current services utilizing location information of wireless nodes, including those above, do not have autonomous mechanisms to exclude illegitimate location information without direct surveillance of the nodes by trusted parties. 

Recently, several authors have proposed an innovative mechanism for wireless secure positioning where all illegitimate location information can be excluded automatically \cite{Waters and Felten} \cite{Capkun and Hubaux}. Their mechanism is based on two fundamental facts: The first is the geometric relation of distances of a point contained within a triangle to triangle's vertices (for 2D planar positioning), or the geometric relation of distances of a point contained within a tetrahedron to tetrahedron's vertices (for 3D spatial positioning). The second is the postulate of modern physics that a propagation speed of information never exceeds the velocity of light. In particular, the authors utilize the measurements of the round-trip time of radio signal propagation and bidirectional communication for variants of the challenge-and-response. 

The proposal in \cite{Waters and Felten}, however, has a vulnerability to the replay attack in the man-in-the-middle scenario, and needs a minor modification to prevent the immediate rebinding of the used session in the challenge-and-response to the false round-trip latency. Meanwhile, the proposal in \cite{Capkun and Hubaux} realizes wireless secure positioning by incorporating the {\it distance bounding protocol} introduced in \cite{Brands and Chaum} into a proposed verification technique called {\it Verifiable Multilateration}. 

In this paper, we propose a novel means to construct the above mechanism by means of unidirectional communication instead of bidirectional communication used in the previous proposals. We assume that a mobile terminal incorporates a verification module as a verifier and a high-precision inner clock in a tamper-resistant protected area, and that the module and the inner clock are protected even from a mobile terminal user. In our proposal, the mobile terminal uses its inner clock and the time and location information broadcasted by radio from trusted stations for positioning.

A similar authentication mechanism for unidirectional communication is found in \cite{Hu Perrig and Johnson} as {\it Temporal Leashes} where both sender and receiver use their tightly synchronized clocks to estimate the traveling distance of radio signals. But the proposal is originally designed to detect the specific attack called the {\it wormhole attack} by checking that the traveling distance of the received packet is below the predetermined upper limit. On the other hand, our protocol does not need any predetermined limit through procedures, but need to include precise location information of the senders into radio signals for the receiver to calculate receiver's own location. Thus, Temporal Leashes \cite{Hu Perrig and Johnson} has significant differences in its purpose and usage from our protocol.

Our proposal has a major advantage in protecting the location privacy of mobile terminal users, because the mobile terminal need not provide any information to the surrounding stations through positioning procedures, thanks to unidirectional communication. Besides, our proposal is free from the positioning error due to claimant's processing-time fluctuations in the challenge-and-response. Our proposal does not need complex key management, and is well-suited for mobile terminals in the open air, or on the move at high speed, in terms of practical usage.

Our proposal depends largely on the advanced hardware technologies such as a high-precision small size clock and a tamper-resistant module. But after examining the present level of clock manufacturing technologies, our hardware requirements are considered feasible and will be materialized in a relatively short period of time, though they are still challenging at this moment.

The organization of this paper is the following. In Section 2, we propose our protocol for wireless secure positioning. In Section 3, we analyze the security of our proposal in comparison to the previous proposals. In Section 4, we discuss functional advantages of our protocol. In Section 5, we discuss the feasibility of our proposal. In Section 6, we review related works. We conclude this paper in Section 7.

\section{Protocol Description}

We propose our protocol for secure positioning on the two dimensional plane as Fig. \ref{fig:proposal}. A digital signature for authentication in Fig. \ref{fig:proposal} can be replaced with a message authentication code (MAC), but an additional measure is necessary for secure secret-key distribution between a verification module and stations.

\begin{figure}[htbp]
\begin{minipage}{\linewidth}
\begin{shadebox}
\begin{minipage}{0.97\linewidth}
\vspace*{2mm}
\begin{enumerate}
\item
A trusted station $\mathsf{S_i}\ (i=1, 2, 3, \ldots)$ computes a digital signature $\mbox{sign}_{s_i}(t_{s_i}, \mbox{\boldmath$x$}_{s_i})$ with $\mathsf{S_i}$'s private key for the future broadcasting time $t_{s_i}$ and $\mathsf{S_i}$'s location $\mbox{\boldmath$x$}_{s_i}$ at the time $t_{s_i}$.
\item
$\mathsf{S_i}$ broadcasts $t_{s_i}, \mbox{\boldmath$x$}_{s_i}$, and $\mbox{sign}_{s_i}(t_{s_i}, \mbox{\boldmath$x$}_{s_i})$ by radio at the time $t_{s_i}$.
\item
A tamper-resistant verification module $\mathsf{M}$ in a mobile terminal receives broadcasts $t_{s_i}, \mbox{\boldmath$x$}_{s_i}, \mbox{sign}_{s_i}(t_{s_i}, \mbox{\boldmath$x$}_{s_i})\ (i=1, 2, 3, \ldots)$ all at once, and at the same time, obtains the time of receipt $t_m$ from the tamper-resistant inner clock. If the number of received broadcasts is less than three, $\mathsf{M}$ aborts the protocol.
\item
$\mathsf{M}$ checks that $t_{s_i} \leq t_m$ for all $i$ of the received broadcasts. If the result is false, $\mathsf{M}$ aborts the protocol.
\item
$\mathsf{M}$ verifies $\mbox{sign}_{s_i}(t_{s_i}, \mbox{\boldmath$x$}_{s_i})$ with $\mathsf{S_i}$'s authentic public key. If the result is true, $\mathsf{M}$ accepts $t_{s_i}, \mbox{\boldmath$x$}_{s_i}$. Otherwise, $\mathsf{M}$ rejects them. If the number of accepted broadcasts is less than three, $\mathsf{M}$ aborts the protocol.
\item
With $t_m$ and all accepted $t_{s_i}, \mbox{\boldmath$x$}_{s_i}$, $\mathsf{M}$ computes $\mathsf{M}$'s location $\mbox{\boldmath$x$}_m$ by applying an appropriate optimization method to triangulation, and also estimates $\mbox{\boldmath$x$}_m$'s error range.
\item
$\mathsf{M}$ checks that $\mbox{\boldmath$x$}_m$'s error range is within the preset limit. If the result is false, $\mathsf{M}$ rejects $\mbox{\boldmath$x$}_m$.
\item
$\mathsf{M}$ verifies that there exists a set of three accepted $\mbox{\boldmath$x$}_{s_i}$ which forms a triangle containing $\mbox{\boldmath$x$}_m$.
If the result is true, $\mathsf{M}$ accepts $\mbox{\boldmath$x$}_m$. Otherwise, $\mathsf{M}$ rejects it.
\end{enumerate}
\vspace{0mm}
\end{minipage}
\end{shadebox}
\end{minipage}
\caption{\ 2D planar positioning protocol with simplex radio communication}
\label{fig:proposal}
\end{figure}

We can easily modify the 2D planar positioning protocol in Fig. \ref{fig:proposal} for secure 3D positioning in a similar fashion to \cite{Waters and Felten} \cite{Capkun and Hubaux}. In the modified 3D positioning protocol, we need at least four, not three, valid time and location information broadcasted by radio from trusted stations, where the validity of each broadcasted information is verified with the time of receipt, an appended digital signature, and station's authentic public key.
In the final domain verification step, the module verifies that receiver's location computed by triangulation is contained within a tetrahedron, instead of a triangle, spatially formed by four trusted stations.

\section{Security Analysis}

\subsection{Distance Bounding}

\begin{figure}[htbp]
\begin{center}
\includegraphics[width=\linewidth, trim=0mm 5mm 0mm 0mm]{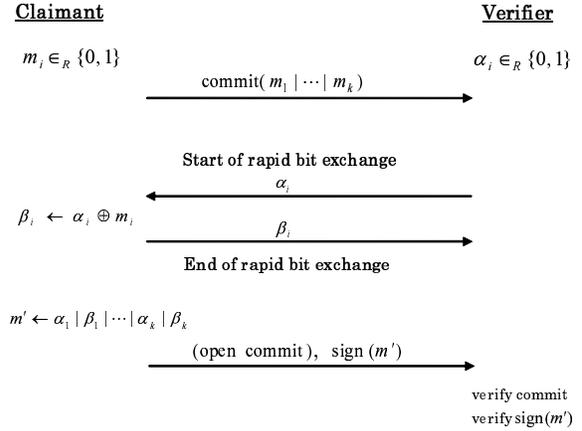}
\end{center}
\caption{Distance bounding with bidirectional communication \cite{Brands and Chaum}. $m_i$ is claimant's nonce and $\alpha_i$ is verifier's nonce. The verifier estimates the upper bound of a distance to the claimant with the round-trip time of rapid bit exchanges. After the rapid bit exchanges, the claimant signs a concatenation of $\alpha_i$ and $\beta_i$ for all $i$ with his private key, and sends it to the verifier.}
\label{fig:duplex_db}
\end{figure}

\begin{figure}[htbp]
\begin{center}
\includegraphics[width=\linewidth, trim=0mm 5mm 0mm 0mm]{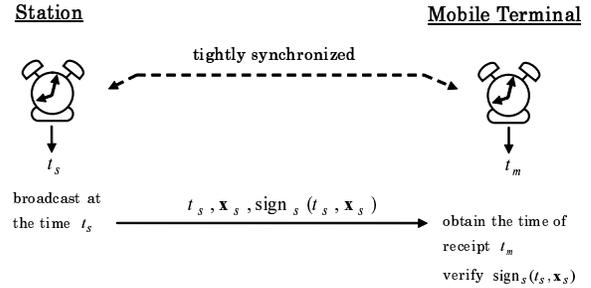}
\end{center}
\caption{Distance bounding with unidirectional communication. The station signs the sending time $t_s$ and station's location $\mbox{\boldmath$x$}_s$ with his private key, and broadcasts them. The mobile terminal estimates the upper bound of a distance to the station with $t_s$ and the time of receipt $t_m$. The mobile terminal then computes its location with three sets of the distance upper bound and $\mbox{\boldmath$x$}_s$.}
\label{fig:simplex_db}
\end{figure}

We will discuss the security of our protocol in terms of the distance bounding in comparison to the previous proposals in \cite{Waters and Felten} \cite{Capkun and Hubaux}.

Distance bounding protocols are first introduced in \cite{Brands and Chaum}, which technically guarantees the distance upper-bound of a device to a verifier by the measurement of the round-trip time of the radio signal propagation. The protocols \cite{Brands and Chaum} are based on the fact that by the forced delay of the radio propagation, an adversary in the man-in-the-middle attack can make a device look further away from a verifier than it truly exists, but cannot make it look closer in principle because no information can propagate faster than light. 

Other than the simple forced delay attack, we will consider two major attacks for the security of the distance bounding below. 
In Fig. \ref{fig:duplex_db} and \ref{fig:simplex_db}, we see the distance bounding protocol with duplex radio communication \cite{Brands and Chaum} and the one with simplex radio communication in our protocol. 
Note that a verifier indicates a trusted station in the original protocol \cite{Brands and Chaum} and the previous proposal \cite{Capkun and Hubaux}, whereas in our proposal a verifier indicates a tamper-resistant module with an inner clock incorporated into a mobile terminal.

\subsubsection{Impersonation Attack}

In this subsection, we analyze the security of distance bounding protocols under the impersonation attack. We define the impersonation attack as a situation where an adversary impersonates another by falsifying authentication information to fool a certain legitimate entity. More specifically, we consider a situation where the adversary falsifies necessary raw data to compute mobile terminal's time and location, and tries to make the verifier believe the false time and location information as true. The adversary may be a malicious third party, or a malicious user of the mobile terminal.

In the case of \cite{Brands and Chaum} \cite{Capkun and Hubaux}, if the adversary falsifies mobile terminal's location by the fake challenge-and-response in the man-in-the-middle attack, the verifier can detect the attack by verifying the committed random number and the submitted digital signature in the last step. The detection succeeds with overwhelming probability as long as the adopted signature scheme is secure and claimant's committed random number is kept secret from the adversary until the challenge-and-response between the adversary and the verifier is completed.

In case of our proposal, if the adversary makes up the arbitrary time and location information, the verifier can detect the attack by verifying the appended digital signature. The detection succeeds with overwhelming probability as long as the adopted signature scheme is secure.

\subsubsection{Replay Attack}

In this subsection, we analyze the security of distance bounding protocols under the replay attack. We define the replay attack as a situation where an adversary repeats the past valid information to deceive a certain legitimate entity. More specifically, we consider a situation where an adversary eavesdrops on the valid communication between a trusted station and a mobile terminal, and fraudulently reuses the past communication to convince the verifier of the validity of the false time and location information. The adversary may be a malicious third party, or a malicious user of the mobile terminal.

In \cite{Brands and Chaum} \cite{Capkun and Hubaux}, if the adversary fraudulently reuses the past valid exchange between a claimant and a verifier, the verifier can detect the attack by verifying the submitted digital signature in the last step. The success probability of the detection is overwhelmingly high as long as verifier's random number is renewed in each challenge-and-response. 

Even if the adversary fraudulently reuses the past valid communication, our proposal still upper bounds the distance of the mobile terminal to the station. Because the time of receipt issued by the tamper-proof inner clock is necessarily later than the past valid time of receipt, the adversary can only lengthen the estimated distance of the mobile terminal to the station, but cannot shorten it.

The proposal in \cite{Waters and Felten} has a vulnerability to the replay attack in the man-in-the-middle scenario. In the attack, an adversary adjacent to Location Manager eavesdrops on the message from Device to Location Manager containing nonces and encrypted Device's ID, and also eavesdrops on the exchanged nonces between Device and Location Manager. The adversary then blocks the valid message from Location Manager to Device, and immediately resends to Location Manager the first used message from Device to Location Manager. Since this time the adversary knows the valid nonces he will exchange to Location Manager in advance, he can fool Location Manager with the false round-trip latency shorter than the valid one, and he can also fool Device with the signed message from Location Manager containing the false round-trip latency. To prevent this attack, Location Manager must check that the encrypted (and randomized) Device's ID in the first message from Device to Location Manager is different from all previously used ones.

\subsection{Positioning with Distance Bounding}

By using the distance bounding in the previous subsection, we construct a secure positioning mechanism in line with the previous proposals of \cite{Waters and Felten} \cite{Capkun and Hubaux}. 

After computing mobile terminal's location by triangulation with distances estimated by the distance bounding, we can verify the validity of the computed location by checking whether the location is inside a triangle formed by trusted stations, or not. As a general geometric property, it holds true that on the plane no point can move to any other point inside a given triangle without shortening any distance of the point to triangle's vertices. 
Since the distance bounding protocol upper bounds the distances of the mobile terminal to the trusted stations, the above geometrical property reliably prevents the malicious adversary from modifying the true location information inside the triangle. 
The same geometric property holds true in the relation between a point and a tetrahedron, instead of a triangle, in 3D space.

In addition to the area test, we check that the error range of the computed location is below the allowable level. With all these filtering tests, we extract only valid positioning results.

But there is a powerful attack to the proposed positioning scheme with the help of plural wireless terminals. We will discuss the details below.

\subsubsection{Collusion Attack}

We define the collusion attack as a situation where plural adversaries share their individual information, or make use of their individual advantages in a cooperative manner, in order to deceive a legitimate entity. In particular, we consider a situation similar to the one discussed in \cite{Capkun and Hubaux} where the adversary colludes with plural wireless terminals placed adjacent to the surrounding trusted stations. 

In case of the distance bounding by the challenge-and-response, each colluded wireless node intercepts the radio signal from the nearest station as a verifier, keeps it for an appropriate length of time, and returns it to the station to make the verifier believe adversary's false distance to the station as true. As discussed in \cite{Capkun and Hubaux}, the previous proposal is secure as long as the verifier can distinguish adversary's colluded wireless nodes from the adversary himself, which means that the secret keys for message authentication codes (MACs) (or the private keys for digital signatures), and the random nonces for the challenge-and-response must be securely protected from the adversary.

\begin{figure}[htbp]
\begin{center}
\includegraphics[width=\linewidth, trim=0mm 30mm 0mm 0mm]{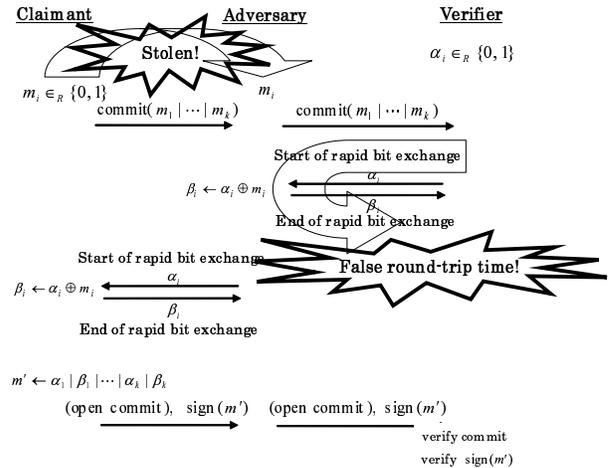}
\end{center}
\caption{Replay attack using a stolen nonce on the distance bounding with bidirectional communication \cite{Singelee}. The adversary makes up the false round-trip time with the stolen nonce. The adversary successfully deceives the verifier without claimant's private key for digital signatures.}
\label{fig:terrorist_attack}
\end{figure}

A variant of the replay attack \cite{Singelee} is shown in Fig. \ref{fig:terrorist_attack}. As pointed out in \cite{Singelee}, if the adversary ever obtains the valid nonces beforehand, adversary's wireless node adjacent to the station uses the stolen nonce for the challenge-and-response with the station (verifier) to make the verifier measure the false round-trip time, reuses verifier's nonce of the previous challenge-and-response for the next challenge-and-response with the mobile terminal (claimant), and results in successful distance falsification by relaying the valid digital signature from the claimant to the verifier. By this means, the colluded wireless nodes can adjust their estimated distances to the nearest station at will in order to look consistent with adversary's false location even inside the verification triangle.

To prevent this type of colluded man-in-the-middle attack, a mobile terminal must securely protect a random number generator for nonces as well as the secret keys (or the private keys) for authentication in the tamper-resistant area.

In case of our proposed protocol, even if adversary's wireless nodes intercept the radio signal on the way to the verification module (verifier) of the mobile terminal, all they can do is to lengthen the estimated distances from the stations but not shorten them, which results in either the computed location of the mobile terminal outside the verification triangle, or the computed location of the mobile terminal inside the verification triangle with the prohibitively enlarged error range. Our proposal is therefore secure against this type of attack as long as the mobile terminal protects its inner clock in the tamper-resistant area.

\section{Functional Advantages}

\subsection{Location Privacy}

In our proposal, a mobile terminal need not provide any information to verify calculated locations because only unidirectional communication is necessary for the verification. This gives mobile terminal users a considerable advantage in protecting their location privacy from the external adversary or the trusted stations. 

In contrast, a protocol mainly discussed in \cite{Capkun and Hubaux} uses bidirectional communication with trusted stations for authentication, where stations learn how far the mobile terminal is located in each verification procedure and a central authority checks the validity of those information. This implies that the mobile terminal users are monitored by the surrounding stations and the central authority through positioning procedures, and the protocol has intrinsic difficulty in protecting the location privacy of the mobile terminal users. 

In \cite{Capkun and Hubaux}, another protocol designed to protect the location privacy is also proposed, where the mobile terminal (or, a tamper-resistant module embedded in the mobile terminal) as a verifier computes and verifies its location with distance bounding to the stations. But it is technically possible for the surrounding stations to extract considerably accurate location information of the mobile terminal from physical properties of the received radio signals, e.g., propagation directions, strengths, or temporal variations, through their bidirectional communication.

\subsection{Positioning Accuracy}

In the previous protocols \cite{Waters and Felten} \cite{Capkun and Hubaux}, positioning is based on the measurements of the round-trip latency of the radio signal from the stations to the mobile terminal. The measured time necessarily includes the processing time to prepare the valid response by the mobile terminal as a claimant, and the uncertainty of the processing time causes considerable positioning error. 

Our protocol with one-way communication is free from the positioning error caused by the  unpredictable processing-time fluctuations mentioned above, because the measured propagation time does not include any intermediate processing time. 

Additionally, in the previous protocols using the challenge-and-response, it is difficult for stations to measure their distances to the mobile terminal on the move. Because the receipt time of stations' challenges tends to be dispersed when the mobile terminal is moving, the location of mobile terminal at the time of distance estimation is easily blurred, which considerably lowers the positioning accuracy. 
Moreover, if received plural challenges from stations must be processed sequentially to prepare valid responses by the mobile terminal, it is difficult to predict the total processing time of the mobile terminal whose fluctuations also cause positioning errors.

In our protocol, after the mobile terminal successfully receives plural broadcasts from stations and the precise time of receipt from the inner clock all at once, the mobile terminal does not have to hurry for positioning accuracy in the subsequent procedures. Today, there are various mobile terminals including GPS receivers which can receive plural broadcasts simultaneously.

\subsection{Coverage Area}

As for radio communication, the size of the coverage area depends largely on the intensity of the transmitted radio wave. In bidirectional radio communication, the size of the coverage area is severely limited by the poor output power of mobile terminal's battery. Since it is technically difficult to increase the output power of small size batteries, the size of the coverage area for bidirectional radio communication cannot be enlarged easily. Hence, if we hope to cover a large area or outdoors for secure positioning, we need considerable number of trusted stations. 

In contrast, our proposal does not have the above-mentioned upper limit of the communication distance, because we only use unidirectional communication from stations to a mobile terminal. If stations have affluent power supplies to send a strong radio wave, the size of the coverage area becomes much larger, and the number of necessary trusted stations becomes much less than the that of previous proposals using bidirectional communication.

\subsection{Key Management} 

As for the mainly discussed proposal in \cite{Capkun and Hubaux}, the setting needs a central authority and a secure backyard network to link between the authority and the stations, where the authority gathers distance information from the stations to compute and verify the mobile user's location. This means the setting needs an additional secure key distribution mechanism to maintain the backyard network, and rather complex key management for it, as shown in \cite{Capkun and Hubaux}. On the contrary, our proposal does not need either a central authority or a secure backyard network for verification of computed locations, because in our setting a verification module in the mobile terminal computes and verifies mobile terminal's position by itself. Thus, our proposal also has an advantage in simple key management.

\section{Feasibility Analysis}

In our proposal, we assume that a verification module and a small size inner clock with high-precision are embedded in the tamper-resistant area of a mobile terminal, and that they are rigorously protected from outside entities including a mobile terminal user. 
We suppose that a mobile terminal is lent by an authority to a user. An inner clock in the mobile terminal is kept isolated from the authority by the time of expiration when the mobile terminal is returned to the authority. The authority checks that the mobile terminal has no irregularities and updates its inner clock. 

In the above usage model, the required precision of the inner clock is roughly approximated with the relation
\begin{equation}
c \times (\delta t \times T) \sim \delta l \label{eq:precision}\ , 
\end{equation}
where $c$ is the velocity of light, $\delta t$ is the precision of the inner clock, i.e., the spontaneous time error of the inner clock, $T$ is the period of validity of the inner clock, and $\delta l$ is the accumulated positioning error due to $\delta t$.
Given that $T=30$ day with the constant $c=3 \times 10^{8}$ m/s, if we want the accumulated positioning error $\delta l$ in the order of $10^{0}$ m, the required precision of the inner clock $\delta t$ should be in the order of $10^{-10}$ s/day.

In fact, several clock manufacturers have already developed small size oven-controlled crystal oscillators (OCXOs) which narrowly meet the above precision requirement with low power consumption \cite{Oscillo} \cite{Vectron} \cite{VF} \cite{C-MAC}. Their typical long term stability is $5 \times 10^{-10}$ s/day, and the accumulated positioning error is about $5$ m according to eq.(\ref{eq:precision}).
Those miniaturized OCXOs are small enough to be incorporated into various types of mobile terminals, and are now available on the market. But the OCXOs are rather sensitive to an abnormal environment and external noises, such as mechanical fluctuations and the high/low temperature, which might be a restriction on some special usages of mobile terminals.

The chip-scale atomic clock developed by NIST is another promising candidate for the inner clock \cite{Knappe1} \cite{Kitching} \cite{Knappe2}. The size of the main unit itself is small enough to be integrated with RFID tags, and even the present size of the total system including surrounding electrical control devices is small enough to be embedded in most mobile terminals \cite{Kitching}. Although the first reported clock precision (of order $10^{-8}$ s/day) \cite{Knappe1} fell short of our requirement above, last year NIST achieved $5 \times 10^{-11}$ s/day for the long term stability \cite{Knappe2}, i.e., about $0.5$ m for the accumulated positioning error by eq.(\ref{eq:precision}), which sufficiently meets our requirement. In addition to the above advantage, the chip-scale atomic clock operates with low power consumption, and is originally designed for low-cost mass production.

Contrary to cryptographic algorithms and techniques, most tamper-resistant hardware techniques have been kept secret among developers exclusively, and there are only limited number of technical literatures available to the public \cite{Anderson and Kuhn}. One well-known measure for tamper-proofing is to set up a trap to certainly detect unauthorized operations or intrusions against the protected area. The detection of the attack immediately triggers to delete the secret data or break the related hardware functions. The mechanism may utilize electrical treatments, irreversible chemical reactions, or mechanical destruction. Another well-known measure for tamper-proofing is to produce the protected area with single-chip integration to cut off the direct contact from outside. 

In our proposal, breaking the inner clock itself cannot be a sufficient countermeasure, because the attacker can freely replace the broken one with his own high-precision clock. We must either forcibly halt the function of the verification module or delete secret identification information as a valid verifier embedded in the protected area. If we choose the chip-scale atomic clock as an inner clock, single-chip integration might be an effective countermeasure.

\section{Related Work}

The secure positioning technique with RF mainly discussed in this paper was proposed in \cite{Waters and Felten} \cite{Capkun and Hubaux}. The distance bounding protocols using bidirectional communication to upper bound claimant's distance was first introduced in \cite{Brands and Chaum}, and the proposal in \cite{Capkun and Hubaux} is based on the protocols \cite{Brands and Chaum}. For easier implementation, a secure positioning technique with a distance bounding protocol using ultrasound and radio communication was proposed in \cite{Sastry}, but it has a security vulnerability to the replay attack due to its use of ultrasound. 
In \cite{Kuhn}, a distance bounding protocol for RFID is proposed. The protocol uses duplex radio communication, and is designed to lessen the processing load of RFID as far as possible.

The protocol called Temporal Leashes is proposed in \cite{Hu Perrig and Johnson} for detection of the specific attack called the wormhole attack. The protocol detects the attack by checking the packet transmission time measured by tightly synchronized clocks of a sender and a receiver.

On the other hand, there are location verification protocols which substantially make use of the physical properties of broadcasted radio waves \cite{Vora} \cite{Singelee}.
In \cite{Vora}, their proposal depends on the intensity and the directivity of broadcasted radio waves for location verification. In \cite{Singelee}, their proposal with duplex radio communication assumes spatial isotropic propagation of radio waves by use of mobile terminal's omni-directional antenna, and uses its particular geometric relation for location verification. But both proposals have a security vulnerability to malicious modification of the assumed physical properties of radio waves. There are many possible ways, especially for a mobile terminal user, to carry out the physical modification of radio waves, e.g., by fraudulently using a directional antenna for the mobile terminal, or by surrounding the mobile terminal with carefully chosen mediums or materials.

\section{Conclusion}

In this paper, we have proposed a novel secure positioning by use of radio broadcasts as unidirectional communication. Our proposal is secure as long as a tamper-resistant module with an inner clock is securely protected. Our proposal has advantages in protecting the location privacy of mobile terminal users, improving positioning accuracy, reducing the number of trusted stations for a large coverage area, and simplifying key management. On the other hand, our proposal depends largely on hardware technologies for a tamper-resistant module and a small size inner clock with high-precision. But we believe those requirements are not serious restriction on our proposal, when considering the consecutive advent of various small size clocks with high-precision. 

In the previous proposals \cite{Waters and Felten} \cite{Capkun and Hubaux}, a random number generator for nonces as well as secret keys for encryption and authentication must be protected even from a mobile terminal user by a tamper-resistant hardware embedded in the mobile terminal. 
In our proposal, correspondingly, a high-precision inner clock must be protected even from  a mobile terminal user by a tamper-resistant hardware embedded in the mobile terminal.

In the near future, our proposal might be useful for an autonomous RFID tag integrated with a micro processor, a small size battery, and a small size high-precision inner clock, which might play a key role to guarantee the traceability in wireless networks.

\section*{Acknowledgements}

The author is profoundly grateful to Koji Yura and Ayumu Shimizu for illuminating and constructive discussions on the subject of this paper. The author is also grateful to Koji Okada and Fumihiko Sano for countless instructive helps in extensive background knowledge.

\end{document}